\documentstyle[twoside,epsf]{article}
\def\ibb #1{\leavevmode\hbox{\kern.3em\vrule
     height 1.5ex depth -.1ex width .4pt\kern-.3em\rm#1}}
\def\Ibb #1{ {\rm I\ifmmode\mkern -3.6mu\else\kern -.2em\fi#1}}

\def\prob(#1|#2){{\Ibb P}\left(#1\mid#2\right)}
\def\Cx {{\ibb C}}

\def\Rx {{\Ibb R}}
\def\N {{\Ibb N}}

\catcode`\@=11
\long\def\@makefntext#1{
\protect\noindent \hbox to 3.2pt {\hskip-.9pt
$^{{\eightrm\@thefnmark}}$\hfil}#1\hfill}       

\def\thefootnote{\fnsymbol{footnote}}
\def\@makefnmark{\hbox to 0pt{$^{\@thefnmark}$\hss}}    

\def\ps@myheadings{\let\@mkboth\@gobbletwo
\def\@oddhead{\hbox{}
\rightmark\hfil\eightrm\thepage}
\def\@oddfoot{}\def\@evenhead{\eightrm\thepage\hfil
\leftmark\hbox{}}\def\@evenfoot{}

\def\sectionmark##1{}\def\subsectionmark##1{}}

\oddsidemargin=\evensidemargin
\renewcommand{\thefootnote}{\fnsymbol{footnote}}

\newcounter{sectionc}\newcounter{subsectionc}\newcounter{subsubsectionc}
\renewcommand{\section}[1] {\vspace{12pt}\addtocounter{sectionc}{1}
\setcounter{subsectionc}{0}\setcounter{subsubsectionc}{0}\noindent
    {\tenbf\thesectionc. #1}\par\vspace{5pt}}
\renewcommand{\subsection}[1] {\vspace{12pt}\addtocounter{subsectionc}{1}
\setcounter{subsubsectionc}{0}\noindent
{\bf\thesectionc.\thesubsectionc. {\kern1pt \bfit #1}}\par\vspace{5pt}}
\renewcommand{\subsubsection}[1] {\vspace{12pt}\addtocounter{subsubsectionc}{1}
    \noindent{\tenrm\thesectionc.\thesubsectionc.\thesubsubsectionc.
    {\kern1pt \tenit #1}}\par\vspace{5pt}}
\newcommand{\nonumsection}[1] {\vspace{12pt}\noindent{\tenbf #1}
    \par\vspace{5pt}}

\newcounter{appendixc}
\newcounter{subappendixc}[appendixc]
\newcounter{subsubappendixc}[subappendixc]
\renewcommand{\thesubappendixc}{\Alph{appendixc}.\arabic{subappendixc}}
\renewcommand{\thesubsubappendixc}
    {\Alph{appendixc}.\arabic{subappendixc}.\arabic{subsubappendixc}}

\renewcommand{\appendix}[1] {\vspace{12pt}
        \refstepcounter{appendixc}
        \setcounter{figure}{0}
        \setcounter{table}{0}
        \setcounter{lemma}{0}
        \setcounter{theorem}{0}
        \setcounter{corollary}{0}
        \setcounter{definition}{0}
        \setcounter{equation}{0}
        \renewcommand{\thefigure}{\Alph{appendixc}.\arabic{figure}}
        \renewcommand{\thetable}{\Alph{appendixc}.\arabic{table}}
        \renewcommand{\theappendixc}{\Alph{appendixc}}
        \renewcommand{\thelemma}{\Alph{appendixc}.\arabic{lemma}}
        \renewcommand{\thetheorem}{\Alph{appendixc}.\arabic{theorem}}
        \renewcommand{\thedefinition}{\Alph{appendixc}.\arabic{definition}}
        \renewcommand{\thecorollary}{\Alph{appendixc}.\arabic{corollary}}
        \renewcommand{\theequation}{\Alph{appendixc}.\arabic{equation}}
        \noindent{\tenbf Appendix \theappendixc #1}\par\vspace{5pt}}
\newcommand{\subappendix}[1] {\vspace{12pt}
        \refstepcounter{subappendixc}
        \noindent{\bf Appendix \thesubappendixc. {\kern1pt \bfit #1}}
    \par\vspace{5pt}}
\newcommand{\subsubappendix}[1] {\vspace{12pt}
        \refstepcounter{subsubappendixc}
        \noindent{\rm Appendix \thesubsubappendixc. {\kern1pt \tenit #1}}
    \par\vspace{5pt}}

\newcommand{\textlineskip}{\baselineskip=13pt}
\newcommand{\smalllineskip}{\baselineskip=10pt}

\def\abstracts#1#2#3{{
    \centering{\begin{minipage}{4.5in}\footnotesize\baselineskip=10pt
    \parindent=0pt #1\par
    \parindent=15pt #2\par
    \parindent=15pt #3
    \end{minipage}}\par}}

\def\keywords#1{{
    \centering{\begin{minipage}{4.5in}\footnotesize\baselineskip=10pt
    {\footnotesize\it Keywords}\/: #1
     \end{minipage}}\par}}

\renewenvironment{thebibliography}[1]
        {\frenchspacing
     \ninerm\baselineskip=11pt
         \begin{list}{\arabic{enumi}.}
        {\usecounter{enumi}\setlength{\parsep}{0pt}
     \setlength{\leftmargin 12.7pt}{\rightmargin 0pt}
         \setlength{\itemsep}{0pt} \settowidth
    {\labelwidth}{#1.}\sloppy}}{\end{list}}

\newcounter{itemlistc}
\newcounter{romanlistc}
\newcounter{alphlistc}
\newcounter{arabiclistc}

\newcommand{\fcaption}[1]{
        \refstepcounter{figure}
        \setbox\@tempboxa = \hbox{\footnotesize Fig.~\thefigure. #1}
        \ifdim \wd\@tempboxa > 5in
           {\begin{center}
        \parbox{5in}{\footnotesize\smalllineskip Fig.~\thefigure. #1}
            \end{center}}
        \else
             {\begin{center}
             {\footnotesize Fig.~\thefigure. #1}
              \end{center}}
        \fi}

\newcommand{\tcaption}[1]{
        \refstepcounter{table}
        \setbox\@tempboxa = \hbox{\footnotesize Table~\thetable. #1}
        \ifdim \wd\@tempboxa > 5in
           {\begin{center}
        \parbox{5in}{\footnotesize\smalllineskip Table~\thetable. #1}
            \end{center}}
        \else
             {\begin{center}
             {\footnotesize Table~\thetable. #1}
              \end{center}}
        \fi}

\def\@citex[#1]#2{\if@filesw\immediate\write\@auxout
    {\string\citation{#2}}\fi
\def\@citea{}\@cite{\@for\@citeb:=#2\do
    {\@citea\def\@citea{,}\@ifundefined
    {b@\@citeb}{{\bf ?}\@warning
    {Citation `\@citeb' on page \thepage \space undefined}}
    {\csname b@\@citeb\endcsname}}}{#1}}

\newif\if@cghi
\def\cite{\@cghitrue\@ifnextchar [{\@tempswatrue
    \@citex}{\@tempswafalse\@citex[]}}
\def\citelow{\@cghifalse\@ifnextchar [{\@tempswatrue
    \@citex}{\@tempswafalse\@citex[]}}
\def\@cite#1#2{{$\null^{#1}$\if@tempswa\typeout
    {IJCGA warning: optional citation argument
    ignored: `#2'} \fi}}

\def\pmb#1{\setbox0=\hbox{#1}
    \kern-.025em\copy0\kern-\wd0
    \kern.05em\copy0\kern-\wd0
    \kern-.025em\raise.0433em\box0}

\def\fnt#1#2{\footnotetext{\kern-.3em
    {$^{\mbox{\scriptsize #1}}$}{#2}}}

\def\fpage#1{\begingroup
\voffset=.3in
\thispagestyle{empty}\begin{table}[b]\centerline{\footnotesize #1}
    \end{table}\endgroup}

\def\runninghead#1#2{\pagestyle{myheadings}
\markboth{{\protect\footnotesize\it{\quad #1}}\hfill}
{\hfill{\protect\footnotesize\it{#2\quad}}}}
\font\tenrm=cmr10
\font\tenit=cmti10
\font\tenbf=cmbx10
\font\bfit=cmbxti10 at 10pt
\font\ninerm=cmr9

\font\eightrm=cmr8

\def\FigName{figure}%
\newbox\captionbox
\long\def\@makecaption#1#2{%
  \ifx\FigName\@captype
    \vskip\abovecaptionskip
    \setbox\tempbox\hbox{{\figurecaptionfont #1\hskip1em #2}}
    \ifdim\wd\tempbox< 28pc
    \centerline{\box\tempbox}
    \else
    {\figurecaptionfont #1\hskip1em #2\par}
\fi\else
    \setbox\tempbox\hbox{{\tablecaptionfont #1\hskip1em #2}}
    \ifdim\wd\tempbox< 28pc
    \centerline{\box\tempbox}
    \else
    {\tablecaptionfont #1\hskip1em #2\par}%
    \fi
 \vskip\belowcaptionskip
 \fi}
\InputIfFileExists{psfig.sty}{\typeout{^^Jpsfig.sty inputed...ok}}{\typeout{^^JWarning: psfig.sty could be be found.^^J}}
\InputIfFileExists{epsfsafe.tex}{\typeout{^^Jepsfsafe.tex inputed...ok}}
            {\typeout{^^JWarning: epsfsafe.tex could not be found.^^J}}
\InputIfFileExists{epsfig.sty}{\typeout{^^Jepsfig.sty inputed...ok}}{\typeout{^^JWarning: epsfig.sty could not be found.^^J}}
\InputIfFileExists{epsf.sty}{\typeout{^^Jepsf.sty inputed...ok}}{\typeout{^^JWarning: epsf.sty could not be found.^^J}}%
\def\fps@figure{tbp}
\def\ftype@figure{1}
\def\ext@figure{lof}
\def\fnum@figure{Fig.\ \thefigure}
\def\qed{\hbox{${\vcenter{\vbox{              
   \hrule height 0.4pt\hbox{\vrule width 0.4pt height 6pt
   \kern5pt\vrule width 0.4pt}\hrule height 0.4pt}}}$}}

\renewcommand{\thefootnote}{\fnsymbol{footnote}}  

\begin{document}
\setlength{\textheight}{7.7truein}    

\runninghead{BELL INEQUALITIES AND ENTANGLEMENT}
            {Reinhard F. Werner and Michael M. Wolf}

\normalsize\textlineskip
\thispagestyle{empty}
\setcounter{page}{1}


\vspace*{0.88truein}

\fpage{1}
\centerline{\bf
BELL INEQUALITIES AND
ENTANGLEMENT}\vspace*{0.37truein}\centerline{\footnotesize
REINHARD F. WERNER AND MICHAEL M. WOLF\footnote{Electronic Mail:
r.werner@tu-bs.de and mm.wolf@tu-bs.de}} \vspace*{0.015truein}
\centerline{\footnotesize\it Institute for Mathematical Physics,
TU Braunschweig} \baselineskip=10pt \centerline{\footnotesize\it
Mendelssohnstr. 3, 38102 Braunschweig, Germany}
\vspace*{0.225truein}

\vspace*{0.21truein}
\abstracts{
We discuss general Bell inequalities for bipartite and
multipartite systems, emphasizing the connection with convex
geometry on the mathematical side, and the communication aspects
on the physical side. Known results on families of generalized
Bell inequalities are summarized. We investigate maximal
violations of Bell inequalities as well as states not violating
(certain) Bell inequalities. Finally, we discuss the relation
between Bell inequality violations and entanglement properties
currently discussed in quantum information theory.
 }{}{}

\vspace*{10pt} \keywords{Bell inequalities, entanglement, local
hidden variable theories}


\vspace*{1pt}\textlineskip  
\section{Introduction and historical survey}   
\vspace*{-0.5pt} \noindent

\setcounter{footnote}{0}
\renewcommand{\thefootnote}{\alph{footnote}}

Quantum mechanics was born in a remarkably short period around the
year 1926, when the long period of guessing turned into the
successful building of the theory, starting a golden age of
remarkable discoveries. In a similar way we can put a date to the
beginning of a particular branch of quantum theory, the theory of
entanglement. It is the year 1935, when both, the paper of
Einstein, Podolsky and Rosen (``EPR''\cite{EPR}), and (motivated
by the EPR paper) Schr\"odinger's article\cite{schroed} in which
he coined the term ``verschr\"ankter Zustand'' (entangled state).

Einstein and Schr\"odinger had both made crucial contributions to
the development of quantum theory, yet they were both expressing a
deep dissatisfaction with the ``present situation of quantum
theory'' (Schr\"odinger's title). And both articles were dismissed
by some of the younger generation as the grumblings of old men who
were just not able to follow the new lines of thought. Bohr's
reply\cite{Bohr} to the EPR paper, although little more than a
refusal to accept the problem, was hailed as a conclusive
rebuttal, and almost everybody went back to business. In
hindsight, however, one must admit that Einstein was struggling
with the deepest departure from classical physics contained in
quantum physics: not the discrete ``jumps'' and other such
conspicuous features, but entanglement. And he was remarkably
lucid, even though we may not share his conclusions.

There are perhaps two reasons in the EPR paper itself which led to
its rather delayed impact on the physics community. One was that
the concern of ``completeness'', seemed like something to worry
about in the distant future, for a community buzzing with
successful applications of the theory. The other was that the
example seemed a bit contrived: the state discussed is a highly
singular object, and even now there are papers trying to make
mathematical sense out of it. As a result the perfect correlation
between results at two distant locations, which was a crucial part
of the argument was not even rigorously true. This defect was
overcome in Bohm's version of the argument\cite{Bohm1}, using
spins (``qubits''). The purely meta-theoretical appeal was changed
dramatically by John Bell\cite{Bell64} in 1964, by the observation
that the EPR dilemma could be formulated in the form of
assumptions, which naturally led to a falsifiable prediction. It
is hardly possible to underrate the importance of this discovery,
which made it possible to rule out not just a particular
scientific theory, but the very way scientific theories had been
formulated for centuries.

The history of experimental verifications of the violations of
Bell's inequalities, predicted by quantum theory, is essentially
the story of building efficient sources for entangled systems. A
breakthrough, bringing the first reliable violations of the
inequalities, was Alain Aspect's atomic cascade\cite{Aspect},
which used the then relatively new technology of optical pumping.
The search of good sources of entangled systems has become much
more intensive with the advent of quantum information theory, in
which entanglement is a key resource. The emphasis has thus
shifted from demonstrating entanglement to using it, and
experimental violations of Bell's inequalities are often merely a
first check whether the source is working properly. New sources
(e.g., using parametric down conversion) now admit Bell
experiments in the student lab. Even the infamous ``detection
loophole'' (related to the fact that only a small fraction of the
produced pairs are really detected) is rapidly being closed
now\cite{detloop}.

On the theoretical side, ``violation of Bell's inequalities'' had
become synonymous with ``non-classical correlation'', i.e.,
entanglement. One of the first papers in which finer distinctions
were made was the construction of states with the property
(Ref.\cite{Werner89}, see Sec.4.2. below) that they satisfy all
the usual assumptions leading to the Bell inequalities, but can
still not be generated by a purely classical mechanism (are not
``separable'' in modern terminology). This example pointed out a
gap between the obviously entangled states (violating a Bell
inequality) and the obviously non-entangled ones, which are merely
classical correlated (separable). In 1995 Popescu\cite{Popescu}
(and later\cite{B96}) narrowed this gap considerably by showing
that after local operations and classical communication one could
``distill'' entanglement, leading once again to violations, even
from states not violating any Bell inequality initially. Similar
examples were then constructed by Gisin\cite{Gisinexp} even for
the case of two qubits. To summarize this phase: it became clear
that violations of Bell inequalities, while still a good indicator
for the presence of non-classical correlations by no means capture
all kinds of ``entanglement''.

The natural conjecture in this situation was that ``violation of
some Bell inequality after suitable distillation'' might be
synonymous with entanglement, i.e., distillation should be
possible for every non-separable state. But in 1998 the Horodecki
family\cite{BE} constructed counterexamples, the so-called {\it
bound entangled states} (see Sec.4.3.). Due to a certain property
(the positivity of the partial transpose) these states turned out
to satisfy any of the known Bell inequalities\cite{PPTBell,WW2}.
Up to now, for the bipartite case, it is neither clear whether the
violation of a Bell inequality already implies distillability nor
do we know whether there is any Bell inequality, which is violated
by a state having positive partial transpose. For multipartite
systems, however, the structure of the state space with respect to
entanglement properties is much richer and D\"ur\cite{DuerBell}
recently showed that there exist indeed undistillable multipartite
states violating a Bell inequality.

\vspace{4pt}The purpose of this article is to give a theoretical
review of the derivation of Bell inequalities from classical
assumptions, discuss their quantum violations and to illuminate
relations to entanglement properties and quantum information
theory in general. Moreover, we emphasize the connection with
convex geometry in the appendix.

As nowadays new papers concerning Bell inequalities or closely
related topics are posted on the Los Alamos e-print archive almost
every day this will by no means be an exhaustive discussion. We
will for instance disregard related topics like the Kochen-Specker
theorem, nonlocal hidden variable theories, experimental
implementations, and the ``Bell theorem without
inequalities\cite{AC}'' . However, these restrictions will enable
us to give an otherwise rather self-contained review of {\it Bell
inequalities and entanglement}. Other review like articles and
extensive discussions emphasizing different topics can be found in
Ref.\cite{CSreview,HSreview,higherspins,Merminrev}.

\section{Derivations of the Inequalities}\label{derivation}

There are many derivations of Bell inequalities in the literature.
This may at first be a bit surprising for such a simple
mathematical statement. However, the hard work in such a
derivation is almost never mathematical but conceptual: if we want
to draw far-reaching conclusions ruling out whole classes of
theories, or ways of formulating natural laws, we have to analyze
theories on a very general and abstract level in order to even
state the assumptions of ``Bell's Theorem''.  Naturally, there are
many ways to say what the really essential assumptions are,
depending on philosophical taste and scientific background.

However, in all derivations two types of elements can be
identified%

\vspace*{4pt}
 \begin{center}
\begin{tabular}{c|c}
  \bf locality \        &\ \  \bf classicality \\
  no-signalling \       &\ \  hidden variables\\
  non-contextuality\ \  &\ \ classical logic\\
                        & \ \ joint distributions\\
                        & \ \ counterfactual definiteness\\
                        &\ \ ``realism''\\
\end{tabular} \end{center}
\vspace*{4pt}
 Since Bell's inequalities are found to be violated in Nature
\cite{detloop}, one of these two assumptions needs to be dropped.
Quantum mechanics (in statistical interpretation) chooses
locality, whereas hidden variable theories drop locality in order
to retain a description by classical parameters. In either case,
however, fundamental features of the pre-quantum way of describing
the world are lost.

\subsection{Basic notation}

Bell type inequalities always refer to correlations between two or
more ``parties'' or sites. It is helpful to imagine that the
experiments at the sites are conducted by physicists,
traditionally named Alice and Bob in the bipartite (two party)
case. Each of the parties gets a particle (or ``subsystem'') from
a common source, and makes a measurement on her/his subsystem. The
basic object of the theory are the joint probabilities obtained in
this way. We will denote a typical measured probability by
\begin{equation}\label{probnote}
  \prob(a_2,b_1|A_2,B_1)\;,
\end{equation} where after the vertical bar we write the devices
used, in this case device $A_2$ by Alice and $B_1$ by Bob, and
before the bar we denote the particular outcomes: $a_2$ a possible
outcome of the device $A_2$ and $b_1$ an outcome of $B_1$. For
simplicity, we will assume throughout that only finitely many
outcomes are possible for each measurement. The collection of all
these numbers are the basic raw data, we might call the {\it
correlation table}.

Of course, these data have to satisfy some constraints which
follow already from the probability interpretation:
$\prob(a_2,b_1|A_2,B_1)\geq0$, and  all the probabilities in a
particular setup $(A,B)$ have to add up to $1$:
\begin{equation}\label{probnorm}
  \sum_{a,b}\prob(a,b|A,B)=1\;.
\end{equation}
 An interesting role is played
by the marginals, which we denote by
\begin{equation}\label{probmarg1}
  \prob(a|A,B)=\sum_{b}\prob(a,b|A,B)\;.
\end{equation}
These are the probabilities measured by Alice in a given setup
$(A,B)$. For general correlation tables such marginals might
depend on the whole setup and, in particular, on the device $B$
chosen by Bob. For example, the device $B$ might be a transmitter
with a particular input fed into it, and $A$ might be a receiver.
Then this dependence on $B$ would be precisely what is required
for Alice to `get the message'. Note, however, that this usually
requires some signal-carrying physical system to go from Bob to
Alice, contrary to the basic description of the correlation setup
(``all parties get particles from a common source''). What we
expect in a general correlation experiment (without communication
between the parties) is the following {\it no-signalling
condition}:
 \begin{eqnarray}\label{probmarg2}
  \sum_{b}\prob(a,b|A,B) \equiv \prob(a|A)
  &\quad&\mbox{is independent of }B
\end{eqnarray}
 and similarly for all other sites, in  this case $A$ instead of $B$.

Before coming to the conditions leading to Bell inequalities we
have to clear up two common misunderstandings concerning hidden
variables and nonlocal effects. These two subsections can be
skipped; the formal development continues in Sec.2.4.

\subsection{Hidden variables exist}

``Hidden variables'' have a bad name in the physics community. Yet
the question whether we can understand the observed quantum
randomness as arising from our ignorance of some underlying
classical variables (this is what the term means) is a fundamental
question, which must be addressed seriously if we want to
understand quantum mechanics at all.

There was an early argument by von Neumann\cite{vNeuHidd} proving
the non-existence of such variables. But von Neumann's proof was
making heavy use of the quantum mechanical structure, so it was
really convincing only for those who had already accepted the
conclusion. The unfortunate consequence was perhaps that some
people began to think that constructing a hidden variable theory,
and thereby contradicting the great von Neumann, was somehow a
non-trivial achievement in itself. What we want to show here is
that, quite to the contrary, it is trivial to construct such a
theory. Moreover, this will allow us later to point out more
precisely the price to be paid in all such theories, namely some
kind of non-locality.

The simplest classical structure from which all measured results
can be obtained was already mentioned: it is the collection of
correlation data itself. This is saying little more than that
gathering statistical data is an activity completely within the
domain of classical probability. Thus in this ``theory'', which is
remarkable only for having explanatory power exactly equal to
zero, the {\it hidden variables are the data to be measured}, and
they are hidden in the same way that the future is. The hidden
variable thus contains a complete description of the experimental
setup, i.e., of the devices $(A,B,C,\ldots)$ chosen by all the
parties\footnote{We could also say that there is a separate
probability space to be chosen for each experimental setup,
although we can equivalently put them all in a single probability
space declaring different setups to be independent.}. This feature
marks so-called {\it contextual} or, more simply, ``non-local''
hidden variable theories.

Contextuality is, of course, not always as blatant, especially in
hidden variable theories focusing on dynamical laws (such as the
Bohm/Nelson theory\cite{Bohm2,Nelson} and its
generalizations\cite{WerHidden}), where the ``setups'' are not
apparent, but enter via a description of the measuring devices
inside the theory. By far the most wide-spread hidden variable
theory is the ``individual state'' interpretation of quantum
mechanics, according to which some wave function is somehow
attached to each individual system, and constitutes a ``catalog of
all expectations'' to be measured on the system. Technically, this
is indeed nothing but the description of a hidden variable theory,
although such statements can also be found by the Copenhagen
Masters, who are not usually associated with hidden variable
views.

\subsection{The Ping Pong Ball Test}

This shows that the temptation is very great to use a language,
which is too naively classical. It is especially great when one
has to explain the quantum world to a general public. This is the
only excuse for the amount of confusing explanations one can find.
Here is a simple guideline for spotting many of the misleading
ones.

{\par\narrower\it\noindent Take any explanations of Bell
inequalities or quantum non-locality, and substitute ping pong
balls for every quantum particle in the account. Then, if what the
author is selling as paradoxical still remains true, he/she isn't
telling you anything about quantum mechanics after all.}

Surprisingly many texts fail this test\footnote{A subtle way of
failing the ping pong ball test is to supplement the description
by the statement that the properties of quantum particles (in
contrast to those of ping pong balls or socks) remain
``objectively undecided'' until the measurement is made. Of
course, this merely shifts the burden of explanation to that
rather cryptic phrase.}\ \
 and lead to ``paradoxes'' like: imagine a
box containing a ping pong ball, which can be separated in two
parts, without looking at the ball. The two parts are then shipped
far apart from each other and after opening one box we then know
``instantaneously'' whether the ball is at the distant location or
not. This is true, but hardly paradoxical, and certainly utterly
useless for sending a message.

\subsection{Local hidden variable theories and the CHSH inequality}\label{CHSHderivation}

To formalize the idea of a local hidden variable theory, let us
explicitly introduce a hidden variable $\lambda$ which takes
values in a space $\Lambda$. We assume that the systems sent to
Alice and Bob (and maybe the others) are described by $\lambda$ in
all details necessary to compute their response to any
measurement, or at least to determine the probabilities of all
such responses. Thus for any measuring device $A$ of Alice, and
any possible outcome $a$ of this device we get a response
probability function $\lambda\mapsto \chi_{A}(a,\lambda)$. The
source of the correlation experiment is characterized by the
probabilities with which the different $\lambda$ occur, i.e., by a
probability measure $M$ on $\Lambda$. With these data we can thus
compute all the correlations
 \begin{equation}\label{lhiv2prob}
  \prob(a,b|A,B)
     =\int\!M(d\lambda)\; \chi_A(a,\lambda)\chi_B(b,\lambda)\;.
\end{equation}
We will say that a correlation table {\it allows a local classical
model}, if it can be represented in this form.

But should this not be always the case? After all, we have only
written down, what any probabilist would write anyhow: there is a
random variable $\chi_A$ for each device, and we are looking at
the joint distribution as we should. But a comparison with the
trivial contextual theories shows immediately where the additional
locality assumption is: in principle the response probabilities
$\chi_A(a,\lambda)$ might also depend on the devices $B,\ldots$
chosen at other sites, and by excluding this dependence we have
increased the demands on our model. This is precisely the
modification leading to Bell's inequalities.

The standard example of a Bell inequality is the {\it
Clauser-Horne-Shimony-Holt} (CHSH) inequality\cite{CHSH}, which
refers to correlation experiments with two $\pm 1$ valued
observables on two sites. From the response probabilities
$\chi_A(a,\lambda)$  ($a=\pm1$) we form the mean value of the
random variable $a$ (given the hidden variable $\lambda$):
 \begin{equation}\label{meanv}
     \hat{a}(\lambda)=\chi_A(+1,\lambda)-\chi_A(-1,\lambda)\;,
\end{equation}
 and from these the correlation function
 \begin{equation}\label{Bcorr}
{\cal B}(\lambda)=\frac12
\big[\hat{a}_1(\lambda)\bigl(\hat{b}_1(\lambda)+\hat{b}_2(\lambda)\bigr)
    +\hat{a}_2(\lambda)\bigl(\hat{b}_1(\lambda)-\hat{b}_2(\lambda)\bigr)
    \big]\;.
\end{equation}
 We claim that ${\cal B}$ satisfies the {\it pointwise} inequality
$|{\cal B}(\lambda)|\leq 1$: indeed, the extreme values are
attained, when each $\hat a_i(\lambda),\hat b_i(\lambda)$ is
extremal, i.e., $\pm1$. But then $2{\cal B}(\lambda)$ is an even
integer, and since $2{\cal B}(\lambda)=4$ requires
$\hat{a}_1(\lambda)=\hat{b}_1(\lambda)=\hat{b}_2(\lambda)=
\hat{a}_2(\lambda)= -\hat{b}_2(\lambda)$, a contradiction, we must
have  $2{\cal B}(\lambda)\leq2$.

Since ${\cal B}$ is pointwise bounded, its expectation, the
so-called {\it Bell correlation},
 \begin{equation}\label{betadef1}
 \beta=\int M(d\lambda){\cal B}(\lambda)
 \end{equation}
is also bounded by unity. But $\beta$ can be expressed directly in
terms of the measured correlation table:  Introducing the
expectation values $E(A,B)=\sum_{a,b=\pm 1} a\; b\;
\prob(a,b|A,B)$, the inequality $\vert\beta\vert\leq1$ becomes the
{\it CHSH inequality}:
 \begin{equation}\label{CHSH1}
 \frac12 \Big| E(A_1,B_1)+E(A_1,B_2) +
E(A_2,B_1)-E(A_2,B_2)\Big| \leq 1 .
\end{equation}

Of course, the existence of local models underlying this
inequality is very reminiscent of the no-signalling condition. In
fact, the locality of the classical model is precisely the
condition that the no-signalling property persists even when we
know the value of $\lambda$, or the source has been upgraded to
produce only one $\lambda$. Conversely, given a local classical
model, the no-signalling condition for the experimental
correlation data is merely ``locality property on average''.

\subsection{Deterministic models and classical configurations}

An important motivation for the search for hidden variables was to
restore the determinism of classical physics or, using Einstein's
famous metaphor, to allow God to quit gambling. We can easily
formulate determinism as a requirement for a local classical
model: the knowledge of $\lambda$ should not only allow us to
predict the probabilities of outcomes, but the outcomes themselves
with certainty. Thus we call a local hidden variable model {\it
deterministic}, if the response functions take only the values $0$
and $1$. It turns out, however, that this seemingly much stronger
constraint on the model does not lead to sharper conditions on the
correlation data\cite{Fine}.

The reason is that we can upgrade any non-deterministic model to a
deterministic one. To do this we only need to incorporate the
randomness in the measuring device into the hidden variable.
Mathematically, we replace the hidden variable $\lambda$ by
$\widetilde\lambda=(\lambda,\xi_A,\xi_B)$, where $\xi_A$ and
$\xi_B$ are uniformly distributed random variables on $[0,1]$,
which are independent of each other and of $\lambda$. We then set
\begin{equation}\label{detexp}
  \widetilde\chi_A(a,\widetilde\lambda)
  \equiv \widetilde\chi_A\bigl(a,\;(\lambda,\xi_A,\xi_B)\bigr)
     =\left\lbrace\begin{array}{cc}
        1& \xi_A\leq \chi_A(a,\lambda)\\
        0& \rm otherwise,
        \end{array}\right.
\end{equation}
and similarly for $B$. Obviously, this model is deterministic, and
it is straightforward to check that it produces the same
correlation table as (\ref{lhiv2prob}).

For a fixed value of the hidden variable $\lambda$ the response
function $\chi_A(a,\lambda)$ in a deterministic model takes on the
value one for one outcome $a$ and vanishes for all the others. If
we now consider an $n$-partite system, where each of the parties
has the choice of $m$ $v$-valued observables to be measured, any
of the $nm$ observables thus divides the hidden variable space
$\Lambda$ into $v$ pieces (which do not necessarily have to be
connected). In this way $\Lambda$ is build up of $v^{nm}$ regions
$\Lambda_c$, such that every region is characterized by a single
{\it classical configuration} $c$, i.e., an assignment of one of
the $v$ outcomes to each of the $nm$ observables (for the CHSH
case for instance with $(n,m,v)=(2,2,2)$ there are 16 classical
configurations). This enables us to rewrite Eq.(\ref{lhiv2prob})
in form of a sum (and analogous for more than two sites)
\begin{eqnarray}\label{lhiv3prob}
 \prob(a,b|A,B)
     &=&\sum_c\int_{\Lambda_c}\!M(d\lambda)\;
     \chi_A(a,\lambda)\chi_B(b,\lambda)\nonumber\\
     &=&\sum_c p_c\; \chi_A(a,c)\chi_B(b,c),
\end{eqnarray}
where $p_c=\int_{\Lambda_c}\!M(d\lambda)$ is the probability
corresponding to the classical configuration $c$. Locality is here
expressed in the fact that the assignment of a value to an
observable at one site does not depend on the observables chosen
at other sites.

In Sec.3. and in the appendix we will see that classical
configurations play a crucial role in the construction of Bell
inequalities as the extreme points of the classically accessible
region.

\subsection{Bell's Telephone}

We will call a {\it Bell telephone} any device, which enables
Alice to send messages to Bob using only the correlations in the
particles the two have obtained from a common source. In other
words, this is precisely the device declared impossible by the
no-signalling or locality conditions. In this section we will give
an alternative proof of the CHSH inequalities, which emphasizes
this communication aspect: we assume a rather weak
``classicality'' condition, and show that Bell's telephone will
work, whenever the CHSH inequality is violated. In fact, we show
that the quality of the transmission is directly related to the
Bell correlation.

We will discuss this in the framework of correlation experiments
used in the previous sections. The framework contains no
space-time aspects so we cannot say that the communication would
be ``superluminal'' (for that, see Sec.5.7.). However, since we
are free to move the partners arbitrarily apart, and the effect
has nothing to do with distance, we can make the communication
superluminal if we want.

If we accept the experimental evidence of violations of the
inequalities, and also uphold the causality principle, which
forbids Bell's telephone, then something must be wrong with the
classicality condition, which is the basic assumption of our
proof. What we assume is that Bob has a joint measurement device,
which simultaneously replaces the two devices he uses in the
experiment for the Bell correlation. In this way, the violations
of Bell's inequalities become a direct experimental verification
of a well known feature of quantum mechanics, namely that there
are observables which do not admit a joint measurement. It also
implies the impossibility of other tasks such as cloning (copying)
and teleportation (transmission of quantum information on a
classical channel).

Let us state the basic assumption in the given framework: Bob has
a {\it joint measuring device} $B_1\&B_2$ for his two observables,
which produces pairs of outcomes $(b_1,b_2)$ with probabilities
$p_i(a_i,b_1,b_2)=\prob(a_i,(b_1,b_2)|A_i,B_1\& B_2)$. The
defining characteristic of such a device is that the statistics of
the outcomes is the same as for the single devices $B_1$ and
$B_2$, i.e. for $i=1,2$:
\begin{eqnarray}\label{jointmeas}
\sum_{b_1}p_i(a_i,b_1,b_2)&=&\prob(a_i,b_2|A_i,B_2)\quad\mbox{and}\\
\sum_{b_2}p_i(a_i,b_1,b_2)&=&\prob(a_i,b_1|A_i,B_1).
\end{eqnarray}
Having this kind of device Bob may guess what apparatus Alice has
chosen by simply interpreting a coincidence of his outcomes
$b_1=b_2$ as ``$A_1$'' and suspecting ``$A_2$'' whenever they
differ. If the probability $p_{\mbox{ok}}$ for Bob to be right is
better than chance ($p_{\mbox{ok}}>\frac12$), then the two can
clearly construct a {\it Bell telephone}. This, however,
immediately takes place as soon as the CHSH inequality is violated
since we can estimate
\begin{eqnarray}\label{pok} p_{\mbox{ok}}&=&
{\frac12}\sum_{a_1,b_1,b_2}\left|\frac{b_1+b_2}2\right| |a_1|
p_1(a_1,b_1,b_2)\nonumber\\
&&\quad+{\frac12}\sum_{a_2,b_1,b_2}\left|\frac{b_1-b_2}2\right|
|a_2| p_2(a_2,b_1,b_2)\nonumber\\ &\geq&
{\frac14}\sum_{a_1,b_1,b_2}(b_1+b_2)a_1
p_1(a_1,b_1,b_2)\nonumber\\
&&\quad+{\frac14}\sum_{a_2,b_1,b_2}(b_1-b_2)a_2
p_2(a_2,b_1,b_2)\nonumber\\ &=&\frac\beta2\;.
\end{eqnarray}
Hence the experimental fact that nature allows $\beta>1$ together
with the no-signalling assumption rules out joint measuring
devices. But this also forbids the existence of {\it universal
cloning} and {\it classical teleportation} since these could be
used to construct a joint measuring device\cite{invitation}.

\section{All the Bell inequalities}\label{allthebells}

So far we have only discussed the CHSH inequality as one specific
example of a Bell inequality. However, there is an infinite
hierarchy of such Bell type inequalities, which can basically be
classified by specifying the type of correlation experiments they
deal with. The essential assumption leading to any Bell inequality
is the existence of a {\it local realistic model}, which describes
the outcomes of a certain class of correlation measurements. The
modus operandi for the derivation of a class of Bell inequalities
would therefore be the following: We first fix the type of
correlation measurements we want to deal with -- say we consider
$n$-partite systems, where each of the parties has the choice of
$m$ $v$-valued observables to be measured \footnote{Of course we
are free to require further restrictions, e.g. we might just want
to look at a subset of all possible correlations or restrict to a
special class of observables or systems we want to investigate.}.
Then we consider the space spanned by the entire set of the raw
experimental data, i.e., the $(mv)^n$ probabilities, and ask for
the inequalities, which bound the region that is accessible within
the framework of a local realistic model. Whatever this underlying
model looks like, if only it is ``classical'', i.e. a local hidden
variable model, the accessible region will be contained in a
convex polytope, whose extreme points are the classical
configurations (for the connection to convex geometry see the
appendix). The classical region is thus bounded by a finite albeit
huge number of linear inequalities. These are the natural
generalizations of the original inequality John Bell published in
1964\cite{Bell64}.

Hence we are faced with a whole hierarchy of inequalities. The
task of finding a minimal set of these inequalities, which is
complete in the sense that they are satisfied if and only if the
correlations considered allow a local classical model, is however
closely related to some known hard problems in computational
complexity\cite{Pitowskybook,Pcomplex}. So it is not surprising
that complete solutions only exist either in cases, where
additional symmetries can be exploited\cite{WW2,ZukowskiBrukner},
or for small values of $(n,m,v)$ where we can utilize today's
computing power for a brute force approach. An extensive numerical
search for the cases $(n,m,v)=(3,2,2)$ and $(2,3,2)$ was performed
by Pitowsky and Svozil\cite{PitowskySvozil}. Unfortunately,
however, the result of such a search is typically a list of the
coefficients of thousands of inequalities\cite{PSweb}, from which
generalizable insights cannot easily be extracted.

Various aspects of the hierarchy of Bell inequalities have been
investigated. Garg and Mermin\cite{higherspins} for instance have
resumed the idea of Bell and discussed systems, with maximal
(anti-) correlation, and $v>2$. Gisin\cite{nsettings} investigated
setups with more than two dichotomic observables per site (however
for arbitrary states). Another closely related subject of
interest, which we will however not discuss, is to expose the
non-local (or non-classical) character of nature without making
use of Bell-type inequalities (cf.
Ref.\cite{MerminGHZ,Kaszlikowski}). In the following we will again
briefly discuss the CHSH\cite{CHSH} inequalities as the first
complete set of inequalities for the case $(n,m,v)=(2,2,2)$. Then
we recapitulate the complete characterization for the multipartite
case $(n,2,2)$ \cite{WW2,ZukowskiBrukner}, where additional
symmetries enable us to give a rather exhaustive discussion.

\subsection{The CHSH case as a complete set of inequalities}

The CHSH inequalities are by far the best studied case of Bell
inequalities. In this case, the characterization is complete in
the following sense, first established by Arthur Fine\cite{Fine}.

The following conditions on a correlation table for two parties
with two dichotomic observables each  ($(n,m,v)=(2,2,2)$) are
equivalent:
 \begin{enumerate}
  \item The correlation table allows a {\it
local realistic model} in the sense of Eq.~(\ref{lhiv2prob}).
 \item The {\it CHSH inequalities} hold, i.e., Eq.~(\ref{CHSH1})
holds, also when any observables at one site or outcomes of a any
observable are interchanged.
 \item There exists a {\it joint probability distribution} for the
outcomes of the four observables, which returns the measured
correlation probabilities as marginals.
\end{enumerate}

We have shown above that 1 and 3 are equivalent, and imply 2.
Hence the non-trivial ``completeness'' part is to show that 2
implies the others, which can be done by analyzing the polytope
discussed above\footnote{Fine originally conjectured that the the
CHSH inequalities (written out for all possible choices of
observables) might even be a complete set for $m>2$. However, Garg
and Mermin\cite{GargMerminvsFine} provided a counterexample with
$(n,m,v)=(2,3,2)$.}.

Item 3 points to another interesting generalization of the CHSH
inequalities.  Given a set of probability distributions
$P_1,\ldots,P_k$, there always exists joint distribution that
returns them as marginals, namely the product distribution
$\prod_{i=1}^k P_i$. However, if we fix in addition the joint
distributions $P_{ij}$ for a certain subset of pairs $(i,j)$, best
visualized as a graph, a joint distribution with these marginals
in general no longer exists.  Bell inequalities thus appear as the
obstructions to extending partial joint probability distributions.

\subsection{All multipartite correlation Bell inequalities for two dichotomic observables per
site}\label{multis}

$n$-particle generalizations of the CHSH inequality were first
proposed by Mermin\cite{Merminineq}, and further developed by
Ardehali\cite{Ardehali}, Belinskii and Klyshko\cite{Klyshko} and
others\cite{Roy,Gisin}. In these works the emphasis was to find
just one inequality for every $n$. In this section we give a
complete set, first constructed in
Refs.\cite{WW2,ZukowskiBrukner}.

The data under consideration are the $2^n$  {\it full correlation
functions} of an arbitrary $n$-partite system, with two dichotomic
observables per site. Each of the $2^n$ different experimental
setups is labeled by the choice of observables at each site. We
will parameterize these choices by binary variables
$s_k\in\{0,1\}$ so that $s_k$ indicates the choice of the $\pm 1$
valued observable $A_k(s_k)$ at site $k$. A ``full correlation
function'' is then the expectation of a product
\begin{equation}\label{xidef}
\xi(s)= E\Big(\prod_k A_k(s_k)\Big),
\end{equation}
where the bit string $s=(s_1,\ldots,s_n)$ labels the respective
experimental setup. Hence $\xi(s)$ can be considered as a
component of a vector $\xi$ in the $2^n$ dimensional space spanned
by the experimental data, and any {\it Bell inequality} is
therefore of the form
\begin{equation}\label{xiBell}
\sum_s \beta(s) \xi(s) \leq 1 ,
\end{equation}
where we have normalized the coefficients $\beta(s)$ such that the
maximal classical value is 1 (i.e., for the CHSH case
$\beta=(\frac12,\frac12,\frac12,-\frac12)$). The linear
combination in Eq. (\ref{xiBell}) may also be computed under the
expectation value, so that this inequality can be stated as an
upper bound to the expectation of
\begin{equation}\label{Bellpolynomial}
{\cal B}=\sum_s \beta(s)\prod_{k=1}^n A_k(s_k).
\end{equation}
We will call such expressions {\it Bell polynomials}. They can be
used directly in the quantum case as {\it Bell operators}, where
all variables $A_k(s_k)$ are substituted by operators with $-{\bf
1} \leq A_k(s_k)\leq{\bf 1}$, acting in the Hilbert space of the
$k$-th site, and the product is taken as the tensor
product.\vspace{8pt}

For the construction of a complete set of inequalities it suffices
to consider the extremal cases, i.e., classical configurations,
where each of the observables takes on one of its two values with
certainty. The restriction to full correlation functions, i.e.,
disregarding correlations of less than $n$ sites, then enables us
to exploit the invariance of $\xi$ under swapping the values of
both observables on two sites. It is basically this symmetry,
which leads to the fact that any of the $2^{2^n}$ binary vectors
$f\in \{-1,1\}^{2^n}$ with components $f(r),\ r\in\{0,1\}^n$
corresponds to one Bell inequality via
\begin{equation}\label{FourierBell}
\beta(s)=2^{-n}\sum_r f(r) (-1)^{\langle r,s \rangle},
\end{equation}
where $\langle r,s \rangle =\sum_{i=1}^n r_i s_i$ denotes the
inner product. Moreover, Eq. (\ref{FourierBell}) indeed provides a
set, which is complete in the sense that the considered
correlations allow a local classical model if and only if all
these inequalities are satisfied.

 Surprisingly, these $2^{2^n}$ linear inequalities are equivalent
to a single non-linear inequality\footnote{The possibility of
replacing the set of linear inequalities by a single nonlinear one
was apparently first recognized by Cirelson for the CHSH case in
Ref.\cite{CC2}.}, namely
\begin{equation}\label{nonlinear}
\sum_r|\hat\xi(r)|\leq 1\quad\mbox{with}\quad
\hat\xi(r)=2^{-n}\sum_s\xi(s)(-1)^{\langle r,s\rangle}.
\end{equation}
This is the characterizing inequality of a {\it hyper-octahedron}
in $2^n$ dimensions. Hence the classical accessible region in this
case has surprisingly high symmetry\cite{WW2}, which is
unfortunately not a symmetry inherent to the underlying problem.
One should thus not expect to find an analogous structure for
other cases of $(n,m,v)$.

\section{Quantum states with no violation}

The violation of Bell's inequality was the first mathematically
sharp criterion for entanglement. In this section we describe
cases in which this criterion fails to detect any entanglement,
even though in some of these cases the quantum state may be
``entangled'' according to now current terminology.

\subsection{Separable states}\label{separablestates}

Generally, entanglement is defined in terms of its negation, and a
quantum state is said to be {\it unentangled, separable} or {\it
classically correlated} iff it can be written as a convex
combination of product states.

Let $\rho$ be a density matrix corresponding to a composite
quantum system described on the Hilbert space ${\cal H}={\cal
H}^{(1)}\otimes{\cal H}^{(2)}$. Then by definition a separable
state can be written as
\begin{equation}\label{separable}
\rho = \sum_j p_j \rho_j^{(1)}\otimes\rho_j^{(2)},
\end{equation}
where the positive weights $p_j$ sum up to one and $\rho^{(i)}$
describes a state on ${\cal H}^{(i)}$. The terminology ``classical
correlated'' is justified due to the fact, that the preparation
leading to the correlations can be assumed to be classical in the
following sense. Suppose we have two independent preparing
devices, one for each subsystem, which prepare a certain state
$\rho_j$ depending on some classical input $j$. Then in order to
obtain a state of the form (\ref{separable}) we have just to add a
random generator, which produces numbers $j$ with probability
$p_j$. Combining these three devices thus leads to such a
``separable'' state and the expectation value of two observables
$A^{(1)}, A^{(2)}$ is then given by
\begin{equation}\label{sepcorrelations}
{\rm tr}\big(\rho A^{(1)}\otimes A^{(2)}\big)=\sum_j p_j {\rm
tr}(\rho^{(1)}_j A^{(1)}) {\rm tr} (\rho^{(2)}_j A^{(2)}).
\end{equation}
The correlation thus just depend on the random generator, which
however can be chosen to be a purely classical device\footnote{We
note that classical correlation does not mean that the state has
actually been prepared in the manner described, but only that its
statistical properties can be reproduced by a classical
mechanism.}. Moreover, it is obvious from Eq.
(\ref{sepcorrelations}) that any separable state admits a
description within the framework of a local classical model (as
described in Sec.2.4.) and therefore satisfies any Bell
inequality. The following subsection is concerned with the fact
that the converse however is not true.

\subsection{$U\otimes U$ invariant states}\label{Wernerstates}

They key idea in Ref.\cite{Werner89} in order to circumvent at
least to some extent both difficulties, the construction of a
local classical model and the proof of non-separability, is to
make extensive use of symmetries. The states considered are those
commuting with all unitaries of the form $U\otimes U$ and can be
written as
 \begin{equation}\label{Wernerstate}
    \rho(p)= (1-p)\frac{P_+}{r_+} + p \frac{P_-}{r_-} , \quad 0\leq p
          \leq 1,
\end{equation}
where $P_+$ ($P_-$) is the projector onto the symmetric
(antisymmetric) subspace of $\Cx^d\otimes\Cx^d$ and $r_\pm = {\rm
tr} P_\pm = \frac{d^2\pm d}2$ are the respective dimensions. It
has been shown that states of the form (\ref{Wernerstate}) are
separable iff $p\leq \frac12$ independent of the dimension of the
system.

Now, consider a von Neumann measurement is performed on each of
the two subsystems $(i=1,2)$:
\begin{equation}\label{vN}
A^{(i)}=\sum_\mu \alpha_\mu^{(i)}Q_\mu^{(i)}\quad\mbox{with}\quad
\sum_\mu Q_\mu^{(i)}= {\bf 1},
\end{equation}
where  the $Q_\mu^{(i)}$ are one-dimensional orthogonal
projectors. A description within a local classical model then
would require that there exist a measure $M$ on a probability
space $\Lambda\ni \lambda$ and response functions
$\chi^{(i)}(\mu,\lambda)\geq 0$ (with $\sum_\mu
\chi^{(i)}(\mu,\lambda)=1$) for any observable, such that
\begin{equation}\label{lcmodel}
{\rm tr}\big(\rho Q_\mu^{(1)}\otimes Q_\nu^{(2)}\big) = \int
M(d\lambda) \chi^{(1)}(\mu,\lambda)  \chi^{(2)}(\nu,\lambda).
\end{equation}
For $\Lambda$ being the unit sphere $\{\lambda\in\Cx^d| \
|\lambda|=1\}$ and the choice
\begin{eqnarray}
\chi^{(1)}(\mu,\lambda)&=&\langle \lambda, Q^{(1)}_\mu
\lambda\rangle,\nonumber\\ \chi^{(2)}(\nu,\lambda)&=& \left\{
  \begin{array}{ll}
    1,\ & \langle \lambda, Q^{(1)}_\nu \lambda\rangle
       <  \langle \lambda, Q^{(1)}_\mu \lambda\rangle\
               \forall\mu\neq\nu \nonumber\\
    0,\ & \mbox{else}\nonumber
  \end{array}
 \right.
\end{eqnarray}
it can be shown that Eq.(\ref{lcmodel}) indeed holds for
\begin{equation}\label{pclassicalbound}
p = 1-\frac{d+1}{2d^2},
\end{equation}
which is in any nontrivial case larger than one half and thus
corresponds to an entangled state. For increasing dimension
Eq.(\ref{pclassicalbound}) even approaches 1, which is (within the
family of $U\otimes U$ invariant states) as far removed from the
classically correlated state ($p=\frac12$) as possible.

\subsection{PPT states}\label{PPTstates}

Another class of states for which it has been
shown\cite{PPTBell,WW2} that none of the inequalities discussed in
Sec.3. are violated is the class of states having a positive
partial transpose ({\it PPT states}). Peres\cite{Peresconjecture}
even conjectured that these states in general admit a description
within a local classical model. Note that without additional
assumptions the converse, however, is not true since the state
(\ref{pclassicalbound}) discussed in the previous subsection
admits such a local classical description although it is no PPT
state.

The partial transpose $A^{T_1}$ of an operator $A$ on ${\cal
H}={\cal H}_1\otimes{\cal H}_2$ is defined in terms of its matrix
elements with respect to a given basis by $ \langle k\ell\vert
A^{T_1} \vert mn \rangle
   = \langle m\ell\vert A\vert kn \rangle
$, and $\rho$ is said to be a {\it PPT state} if $\rho^{T_1}\geq
0$.

The key idea for showing that PPT states satisfy any inequality
coming from Eq.(\ref{FourierBell}) is to utilize the positivity of
the partial transpose by applying the variance inequality
\begin{equation}\label{variance} {\rm tr}(\rho{\cal B})^2={\rm tr}(\rho^{T_1}{\cal
B}^{T_1})^2\leq {\rm tr}\big(\rho^{T_1}{{\cal B}^{T_1}}^2\big),
\end{equation}
where $\cal B$ is the {\it Bell operator}, whose expectation value
has to be bounded by one within a local classical model.
Additionally averaging over all partial transposes with respect to
any partition of the multipartite system into two subsystems then
shows that in fact ${\rm tr}(\rho{\cal B})\leq 1$.

Though the positivity of the partial transpose is known to be one
of the most efficient separability criteria\cite{PeresPPT} it is
in general not a sufficient one. Hence there exist states, which
are not classically correlated although they satisfy the PPT
condition and therefore admit a (however possibly restricted)
local classical description\cite{BE}. These states are often
referred to as {\it PPT bound entangled states} since they have
the additional property that their entanglement cannot be
recovered by entanglement distillation\cite{dis}.

\section{Quantum violations: The CHSH case}

\subsection{The See-Saw iteration}
The derivation of the maximal quantum violation of a Bell
inequality for an arbitrary state is a high dimensional
variational problem for which we are not aware of any explicit
solution except for the case of the CHSH inequality for two qubit
systems. Therefore we begin with providing a simple iterative
algorithm, the {\it See-Saw iteration}, which turned out to be an
efficient method for maximizing an affine functional with respect
to hermitian operators with $-{\bf 1}\leq A\leq {\bf 1}$
corresponding to the expectation of a Bell operator with
dichotomic observables. Since generalization to the multipartite
case is straight forward, we content ourselves with bipartite
systems where the functional to be maximized is of the form
\begin{equation}\label{Bellbipartite}
{\rm tr}\big( {\cal B}\rho\big)=\sum_{i,j}{\rm tr}\big( \beta_{ij}
A_i\otimes B_j\rho \big),
\end{equation}
where it suffices to consider
 unitary observables $A=A^*=A^{-1}$ (and the same for $B$), as these are
extremal in the convex set of Hermitian operators with $-{\bf 1
}\leq A\leq{\bf 1}$. The idea is now to maximize this functional
with respect to observables on one site while keeping the other
ones fixed, and then to iterate this procedure. Therefore, we
rewrite Eq.(\ref{Bellbipartite}) by taking the partial trace over
that site of the system, which we keep fixed,
i.e.:\begin{eqnarray} {\rm tr}\big( {\cal B}\rho\big)&=& \sum_i
{\rm tr}_A\big(X_i A_i\big)=\sum_j {\rm tr}_B\big(Y_j
B_j\big)\quad\mbox{with}\label{ff1}\\ && X_i={\rm tr}_B\Big(\sum_j
\beta_{ij}\big({\bf 1}\otimes B_j\big)\rho\Big)\\ && Y_j={\rm
tr}_A\Big(\sum_i \beta_{ij}\big(A_i\otimes{\bf 1}\big)\rho\Big)
\end{eqnarray}
The maximization in Eq.(\ref{ff1}), however, can be made explicit
just by taking $A_i=\mbox{sign}(X_i)$ resp.
$B_j=\mbox{sign}(Y_j)$. Of course, the {\it See-Saw iteration} is
faced with the usual problem of most numerical optimization
methods: it cannot guarantee the convergence on absolute extrema.
Nevertheless it turned out to be a useful tool in the search for
Bell violations (e.g. in Ref.\cite{WW2}), which converges already
after a few iterations and is in general very stable with respect
to variations of the initial values.

\subsection{Cirelson's inequality}\label{Cirelson's inequality}

The best upper bound for the violation of the CHSH inequality,
first derived by Cirelson\cite{Cirelsonbound}, is obtained by
squaring the Bell operator and utilizing the variance
inequality\cite{Landau}, which already appeared in the previous
section in Eq.(\ref{variance}). Taking again into account that it
suffices to consider
 unitary observables of the form $A=A^*=A^{-1}$ we get
\begin{equation}\label{B2}
{\cal B}^2 = {\bf 1} - \frac14 [A_1,A_2]\otimes [B_1,B_2].
\end{equation}
Since the commutators are bounded by two as
$\Vert{[A,B]}\Vert\leq2\Vert A\Vert\;\Vert B\Vert$ this leads to
\begin{equation}\label{Cir}{\rm tr} (\rho {\cal
B})\leq\sqrt{2},\end{equation} which is usually referred to as
{\it Cirelson's inequality}.

\subsection{Operators for maximal violation}

The bound $\Vert{[A,B]}\Vert\leq2\Vert A\Vert\;\Vert B\Vert$ used
in the previous section is clearly saturated when $A$ and $B$ are
Pauli matrices. It is therefore no surprise that experiments
coming close to the maximal violation $\beta\approx\sqrt2$ are
possible with qubit systems, and indeed this is precisely the
idealized description of Aspect's experiments. In this subsection
we argue (following Ref.\cite{SW2}) that the qubit example is even
the {\it only} possibility to get the maximal violation in any
dimension.

To give a more precise, but simple statement, suppose that both
the restricted density operators are faithful (have no zero
eigenvalues), and suppose $A_1,A_2,B_1,B_2$ are operators giving
$\beta=\sqrt2$. Then $A_1,A_2$ and $A_3=i A_2A_1$ satisfy all the
algebraic relations of the Pauli matrices: $A_k^2={\bf 1}$, $A_1
A_2=iA_3$ and cyclic permutations thereof.

The basic idea of the proof is to write the inequalities in terms
of the two operators
 \begin{eqnarray}
 A&=&\frac12(A_1+iA_2)\\
 B&=&\frac1{2\sqrt2}\big[(B_1+B_2)+i(B_1-B_2)\big]
 \end{eqnarray}
which allow a simple representation of the Bell operator as
 ${\cal B}=\sqrt2(A^*B+AB^*)$. One readily checks that, on the other
hand, $A^*A+AA^*\leq\bf1$ and $B^*B+BB^*\leq\bf1$. The core of the
proof is the decomposition
\begin{eqnarray}
 {\bf1}-{\cal B}/\sqrt2
    &=& (A-B)^*(A-B)+(A-B)(A-B)^*+ \nonumber\\
    &&\qquad+({\bf1}-A_1^2)+({\bf1}-A_2^2)
            +({\bf1}-B_1^2)+({\bf1}-B_2^2)\;.
 \end{eqnarray}
 Clearly, the right hand side is a positive operator, which
provides an alternative proof of Cirelson's inequality
(\ref{Cir}). States with maximal violation are those for which
this equation, and hence every single term on the right hand side
has expectation zero. In particular, for any vector $\Phi$ in the
support of the density operator we get
$(A-B)\Phi=(A^*-B^*)\Phi=({\bf1}-A_1^2)\Phi=\cdots=0$. Hence
$(A^2+A^{*2})\Phi=(1/2)(A_1^2-A_2^2)\Phi=0$ and
$(B^2-B^{*2})\Phi=(i/2)(B_1^2-B_2^2)\Phi=0$. Combining this with
$A^2\Phi=B^2\Phi$ and  $A^{*2}\Phi=B^{*2}\Phi$ we get $A^2\Phi=0$
and similarly $(A^*A+AA^*)\Phi=\Phi$ for every vector $\Phi$ in
the support of the density operator. Since the reduced density
operator for Alice has full support, this implies the identities
$A^*A+AA^*=\bf1$ and $A^2=0$, so that $A_1=A+A^*, A_2=i(A^*-A)$,
and $A_3=A^*A-AA^*$ are a realization of the Pauli matrices.

\subsection{Qubits: Structure of the Bell operator}

It is obvious from Eq.(\ref{B2}) that as soon as the observables
on only one of the two subsystems commute, the inequality is
satisfied. We may therefore disregard the case $A={\bf 1}$ and for
the case of two qubits restrict to observables of the form
$A_k(s_k)=\vec{a}_k(s_k)\vec{\sigma}$, where $\vec{\sigma}$ is the
vector of Pauli matrices and $\vec{a}_k(s_k)$ is a normalized
vector in $\Rx^3$. Furthermore, we can use the homomorphism
between $SU(2)$ and $SO(3)$ and do a local unitary transformation
such that the vectors belonging to the four observables all lie in
the $x-y$ plane, i.e.:
\begin{equation}\label{planeobs}
A_k=\sigma_1 \sin\alpha_k + \sigma_2 \cos\alpha_k
\end{equation}
With this choice of observables we get
\begin{equation}\label{B2plane}
{\cal B}^2={\bf 1}
-\sin(\alpha_1-\alpha_1')\sin(\alpha_2-\alpha_2')\sigma_3\otimes\sigma_3,
\end{equation}
such that the four {\it Bell states}
$|\Phi^\pm\rangle=\frac1{\sqrt{2}}(|00\rangle\pm|11\rangle)$ and
$|\Psi^\pm\rangle=\frac1{\sqrt{2}}(|01\rangle\pm|10\rangle)$ turn
out to be eigenstates of the CHSH operator. Moreover, $\cal B$ has
symmetric spectrum since $({\bf 1}\otimes\sigma_3){\cal B}=-{\cal
B}({\bf 1}\otimes\sigma_3)$ so that we end up with a Bell operator
which is up to local unitary transformations always of the
form\cite{Gisindecomposition}
\begin{equation}\label{CHSHspectral}
{\cal B}=\lambda_1
\big(P_{\Phi^+}-P_{\Phi^-}\big)+\lambda_2\big(P_{\Psi^+}-P_{\Psi^-}\big),
\end{equation}
where $P_\cdot$ denotes the projector onto the respective Bell
states and the eigenvalues have to satisfy
$\lambda_1^2+\lambda_2^2=2$, which follows from ${\rm tr}{\cal
B}^2=4$.

In fact, the spectral decomposition in Eq.(\ref{CHSHspectral})
exhibits a property of the Bell operator, which is typical for any
multipartite inequality for two dichotomic observables per site
(see Sec.6.2.).

\subsection{Qubits: Maximal violation for arbitrary states}

Let us now consider an arbitrary quantum state $\rho$ of two
qubits and let $R_{ij}={\rm tr}(\rho \sigma_i\otimes\sigma_j)$.
Following Ref.\cite{HHHCHSH} the maximal violation of the CHSH
inequality is then given by
\begin{eqnarray}\label{CHSHmax1}
\beta(\rho)&=&\max_{\vec{a_1},\vec{a_1'},\vec{a_2},\vec{a_2'}}\frac12\big(\vec{a_1}\cdot
R(\vec{a_2}+\vec{a_2'})+\vec{a_1'}\cdot
R(\vec{a_2}-\vec{a_2'})\big)\nonumber\\
&=&\max_{\vec{a_2},\vec{a_2'}}\frac12\big(||R(\vec{a_2}+\vec{a_2'})||+
               ||R(\vec{a_2}-\vec{a_2'})||\big)\nonumber\\
&=&\max_{\varphi,\vec{c}\perp\vec{c'}}\cos\varphi
||R\vec{c}||+\sin\varphi ||R\vec{c'}||\nonumber\\
&=&\max_{\vec{c}\perp\vec{c'}}\sqrt{||R\vec{c}||^2+||R\vec{c'}||^2},
\end{eqnarray}
where the maxima are always taken over all unit vectors.
Evaluating the last maximum we obtain
\begin{equation}\label{CHSHmax}
\beta(\rho) = \sqrt{\nu+\nu'} ,
\end{equation}
where $\nu, \nu'$ are the two largest eigenvalues of the matrix
$R^TR$. For pure two qubit states, which can always be written in
their Schmidt form as $|\Psi\rangle=\cos\varphi |00\rangle +
\sin\varphi |11\rangle$ this can further be evaluated to
\begin{equation}\label{pureviol}\beta(\Psi)=\sqrt{1+\sin^2(2\varphi)}
 ,\end{equation} which means that as
soon as a pure two qubit state is entangled it violates the CHSH
inequality\cite{Gisinpure}\footnote{This result was generalized to
higher dimensional bipartite systems by Gisin and
Peres\cite{GisinPeres}.}.

\subsection{Continuous variable systems }
There is recent effort in order to adopt the CHSH inequality to
the {\it continuous variable} case\cite{BW98,BW99,JLK00,CPHZ01}.
One possibility in order to derive dichotomic observables in this
case is to utilize the apparent analogy between the parity
operator in Fock space and the ``spin-measurement'' in $\Cx^2$
associated to the Pauli $\sigma_3$ operator\footnote{For a
different approach see for instance Ref.\cite{BW99}, where the
chosen observables distinguish between photon counts and no photon
counts.}. As admissible observables in the continuous variable
case we may then either use the set of coherently displaced parity
operators \cite{BW98} or explicitly construct a direct analogue to
the three Pauli operators\cite{CPHZ01} by establishing the
isomorphism $\ell^2(\N)\simeq\Cx\otimes \ell^2(\N)$ via collecting
parities. Let
\begin{eqnarray}
s_z &=& \sum_{n=0}^\infty\big(|2n\rangle\langle 2n| -
|2n+1\rangle\langle 2n+1|\big) ,\label{s1} \\
s_+&=&\sum_{n=0}^\infty |2n\rangle\langle 2n+1 |,\quad s_-=s_+^*
,\label{s2}
\end{eqnarray}
be the parity and parity flip operators in Fock state
representation, then $(s_x,s_y,s_z)$ with $s_x\pm i s_y=2 s_\pm$
is again a representation of the Pauli matrices. So we can in
principle make use of all the above results and
Eq.(\ref{CHSHmax1},\ref{CHSHmax}) lead to a lower bound on the
maximal violation of the CHSH inequality for continuous variable
states.

A special kind of such states are {\it Gaussian states}, i.e.,
states having a Gaussian Wigner distribution, which play a crucial
role in quantum optics, since coherent, squeezed and thermal
states are all Gaussians. If we only consider observables given by
the field quadratures or simple functions thereof,  then the
positive Wigner function itself provides a ``hidden variable
distribution''. Hence no violation of a Bell inequality can occur.
Observables obtained from Eq.(\ref{s1},\ref{s2}), however, are not
of that kind. For the pure Gaussian two mode state
\begin{equation}\label{Gausspure}
|\psi(r)\rangle = \sum_{n=0}^\infty c_n |n\rangle\otimes
|n\rangle\; ,\quad c_n=\frac{\tanh^n(r)}{\cosh(r)},
\end{equation}
which is characterized by the squeezing parameter $r$, the maximal
violation with respect to these observables is
\begin{equation}\label{gaussbeta}
\beta(r)=\sqrt{1+\tanh^2(2r)}
\end{equation} analogous to Eq.(\ref{pureviol}).

\subsection{Quantum Field Theory}\label{sec:QFT}

It is not surprising that Quantum Field Theory should contain
violations of Bell inequalities. After all, this theory is
supposed to describe Aspects experiment, too. There are three
special reasons, though to at this theory specifically. The first
is that here one can take the locality assumption of the general
derivations literally in the sense of Einstein causality. Thus
Alice and Bob are assigned space time regions in which they can
perform their experiments, and these regions are chosen to be
spacelike separated, so that relativistic causality forbids any
signaling between the two. Since we are looking at a {\it quantum}
field theory, it is clear that quantum features are also contained
in the description from the outset. The best adapted framework for
this discussion is ``algebraic quantum field
theory\cite{HaagKastler}'' also called ``Local Quantum
Physics\cite{Haag}'': here the concepts of quantum structure and
relativistic localization are taken as the axiomatic starting
point.

The second feature making quantum field theory interesting is that
we have here a distinguished state, the vacuum. As it is well
known that there are always vacuum fluctuations, it is natural to
ask whether these fluctuations are classical or not. More
concretely: if Alice and Bob are in spacelike separated
laboratories, can they get a violation of Bell's inequalities from
vacuum fluctuations alone? It turns out\cite{SW1} that they can,
although the effect is extremely small for large spatial
separation: in a massiv theory, it decreases exponentially with
the separation of the regions on the scale of the Compton
wavelength. On the other hand, one gets maximal violation if the
regions are very close\cite{SW2}.

The third reason to look a quantum field theory emerges from these
studies: it turns out that not only the vacuum produces maximal
violations at short range, but {\it any} state, which is not
extremely singular\footnote{Technically speaking: any state which
is locally normal with respect to the vacuum} (e.g., requires only
finite total energy). This is a new possibility arising only in
theories with infinitely many degrees of freedom, and quantum
field theory is an ideal testing ground for ideas about this
phenomenon.

We remark that it is still an open problem to show that even at
arbitrarily large distance a (necessarily exponentially small)
violation of the CHSH inequality can be detected. What is known so
far is only that the positive partial transpose property always
fails\cite{VerchW}, and hence the vacuum is not classical at any
distance.

\section{Quantum Violations: Beyond CHSH}
\subsection{Bipartite systems with more than two observables}

For the case of more than two dichotomic observables per site only
little is known. In particular there is yet no explicit
characterization of the extremal inequalities, although
constructing some inequalities, e.g. by chaining CHSH
inequalities\cite{BraunsteinCaves}, is not difficult. However,
Cirelson\cite{CC1,CC2} recognized that the quantum correlation
functions, which are in general rather cumbersome objects, can be
reexpressed in terms of finite dimensional vectors in Euclidean
space.

If we have two sets of observables $\{A_1(s)\}$ and $\{A_2(t)\}$
(again hermitian with $-{\bf 1}\leq A\leq {\bf 1}$) where
$s\in\{1,\ldots,p\}$ and $t\in\{1,\ldots,q\}$, then for any state
$\rho$ there exist sets of real unit vectors $\{x_s\}$ and
$\{y_t\}$ in the Euclidean space of dimension $q+p$ such that
\begin{equation}\label{Cirelsonvec}
{\rm tr}\big(\rho A_1(s)\otimes A_2(t)\big) = \langle x_s,
y_t\rangle\quad \forall_{s,t}.
\end{equation}
For the case of two observables on one site and an arbitrary
number on the other Cirelson showed that the maximal quantum
violation is $\sqrt{2}$ and thus already obtained for the CHSH
inequality ($p=q=2$). For an increasing number of observables on
both sites, however, he obtained the {\it Grothendieck constant}
($\approx 1.782$), known from the geometry of Banach spaces, as a
limit for the maximal violation.

\subsection{The multipartite case -- the role of GHZ states and Mermin's
inequality}\label{multipartitecase}

Like the four Bell states are of particular importance for the
CHSH operator, the {\it generalized GHZ states}\cite{GHZ89}, which
are up to local unitaries of the form
\begin{equation}\label{GHZ}
|\Psi_{GHZ}\rangle=\frac1{\sqrt{2}}\big(|0 0 \ldots 0\rangle + |1
1 \ldots 1 \rangle\big),
\end{equation}
play a special role for all multipartite inequalities with two
dichotomic observables per site. In fact, they provide a basis of
eigenstates for all the Bell operators with extremal observables
and lead thus to maximal violations\cite{Gisindecomposition,WW2}.

If we again consider observables of the form in
Eq.(\ref{planeobs}), which are obtained after applying local
unitaries, and let $\Omega\in\{0,1\}^n$ and its complement
$\overline{\Omega}$ with $\overline{\Omega}_k=1-\Omega_k$
characterize vectors on the $n$-fold tensor product, then
\begin{eqnarray}\label{Aomega}
\bigotimes_{k=1}^nA_k(s_k)|\Omega\rangle &=&
f_\Omega(s)|\overline{\Omega}\rangle,\quad \mbox{ with }\
\\ f_\Omega(s)&=&\exp\big[i\sum_{k=1}^n
\alpha_k(s_k)(-1)^{\Omega_k}\big].
\end{eqnarray}
Therefore the set of $2^n$ GHZ-like basis vectors
$|\Psi_\Omega\rangle=\frac{1}{\sqrt{2}}(e^{i\theta_\Omega}|\Omega\rangle+|\overline{\Omega}\rangle)$
satisfies the eigenvalue equations ${\cal
B}|\Psi_\Omega\rangle=\lambda_\Omega|\Psi_\Omega\rangle$ if
$\theta_\Omega$ is chosen such that
\begin{equation}\label{lambdatheta}
\lambda_\Omega=e^{i\theta_\Omega}\sum_s\beta(s)f_\Omega(s)
\end{equation} is real. Hence any Bell operator for multipartite
systems with two dichotomic observables per site indeed admits
spectral decomposition into GHZ states\cite{Gisindecomposition},
and it is an immediate corollary thereof that GHZ states lead to
maximal violations. It was shown in Ref.\cite{WW2} that the
computation of the latter can be reduced to a variational problem
with just one free variable per site. Moreover, even any extreme
point of the convex body of the quantum mechanically attainable
correlation functions is found in the generalized GHZ
states\cite{WW2}.

It is crucial for the derivation of all the above results that we
have no more than two dichotomic observables, since three or more
vectors do in general not lie in a plane and we would therefore
not be able to restrict to observables of the form in
Eq.(\ref{planeobs}).

If we now fix the state to be of the GHZ form (\ref{GHZ}) and ask
for the inequality within the set (\ref{FourierBell}) leading to
the maximal violation, we obtain {\it Mermin's inequality}
\cite{Merminineq}\footnote{In fact, Mermin\cite{Merminineq}
derived the inequality corresponding to Eq.(\ref{Bellrecursive})
for odd $n$, whereas Ardehali \cite{Ardehali} in turn obtained the
one for even $n$. It was then Klyshko and Belinskii\cite{Klyshko},
who recognized that Eq.(\ref{Bellrecursive}) covers both types of
inequalities (see also Ref.\cite{Gisin}). However, since the basic
idea is going back to David Mermin it seemed justifiable to us to
refer to the inequality as ``Mermin's inequality''.}. For an
$n$-partite system the Bell operator corresponding to Mermin's
inequality is defined recursively starting with ${\cal B}_1=A_1$
by
 \begin{equation}\label{Bellrecursive}
{\cal B}_n = \frac{{\cal
B}_{n-1}}2\otimes\big(A_n+A_n'\big)+\frac{{\cal
B}_{n-1}'}2\otimes\big(A_n-A_n'\big),
\end{equation}
where ${\cal B}_n'$ is obtained from ${\cal B}_n$ by exchanging
all the contained observables $A\leftrightarrow A'$ \footnote{A
different way of deriving Mermin's inequality is obtained by
identifying the expectation values of observables $A$ and $A'$ of
a single site with a square in the complex plane. After a suitable
linear transformation (a $\pi/4$ rotation and a dilation) we can
take it as the square ${\cal S}$ with corners $\pm 1$ and $\pm i$.
The pair of expectation values of $A$ and $A'$ is thus replaced by
the single complex number ${\rm tr}(\rho a)$, where
 $a={1\over2}\
 \bigl((A+A')+i(A'-A)\bigr)=e^{-i\pi/4}\,(A+iA')/\sqrt2$.
  The basic idea behind this transformation is that products of
complex numbers lying in $\cal S$ again lie  in $\cal S$. Since
this also holds for convex combinations and pure states within a
local classical model are always of a product form, the statement
that the product of such complex expectations lies in $\cal S$
indeed corresponds to a Bell inequality. In fact, this is
essentially Mermin's inequality (see Ref.\cite{PPTBell}).}.
Permitting arbitrary unrelated choice for the $n-1$ partite Bell
operators ${\cal B}_{n-1}$ and ${\cal B}_{n-1}'$ it was shown in
Ref.\cite{WW2} that any inequality of the set (\ref{FourierBell})
can be obtained from Eq.(\ref{Bellrecursive}), i.e., by nesting
CHSH inequalities. Squaring the Bell operator as in Eq.(\ref{B2})
leads to  ${\rm tr}(\rho{\cal B}_n)\leq 2^{\frac{n-1}2}$, which is
however only saturated for Mermin's inequality\cite{WW2}. Hence
the maximal violations grow exponentially with the number of
subsystems. However, one has to keep in mind that the joint
efficiency of $n$ independent detectors would in turn decline
exponentially in $n$.

Finally we should mention that for the case of multipartite
systems with an infinite set of settings, i.e. more than two
dichotomic observables per site,
Zukowski\cite{Zukowskiallsettings} derived an inequality for which
the GHZ state leads to an exponentially growing violation of
$\frac12\big(\frac{\pi}2\big)^n$.

\section{Relations to Quantum Information Theory}

One of the essential innovations of quantum information theory is
to think of entanglement as a resource for quantum information
processing purposes. This new point of view led to a dramatic
increase of knowledge about the structure of the state space with
respect to entanglement properties. Whereas in the late eighties
there was hardly any difference between entangled states and
states violating some Bell inequality, we have a much more subtle
discrimination nowadays.\vspace*{4pt}

\begin{figure}[htbp] 
\vspace*{13pt}
\centerline{\psfig{file=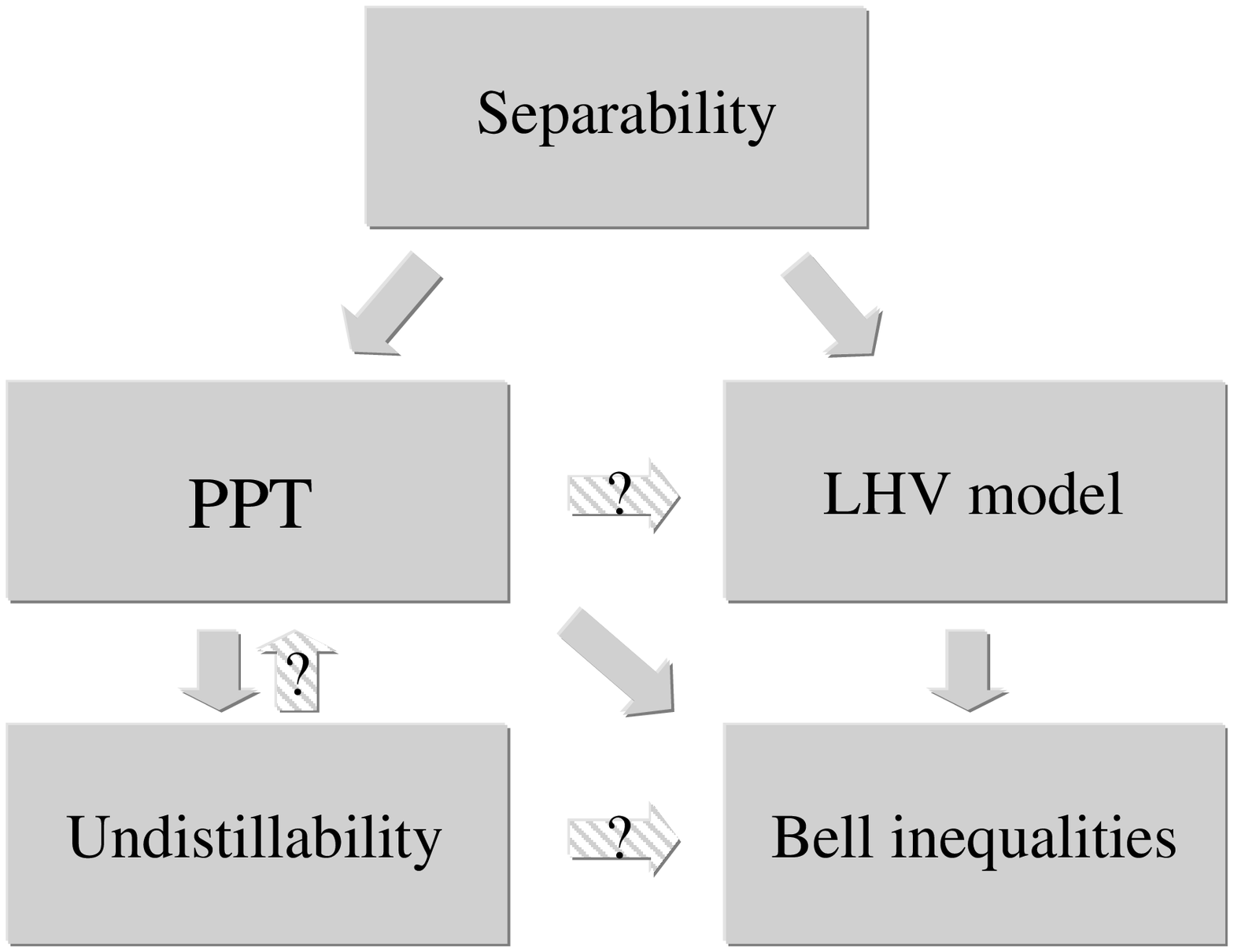,width=8.2cm}} 
\vspace*{13pt} \fcaption{Relations between various degrees of
``classicalness''.\label{fig1}}
\end{figure}

Figure \ref{fig1} summarizes the previously discussed relations
between various degrees of classicalness, i.e., negations of
entanglement properties. The implication ``PPT $\Rightarrow$ Bell
inequalities'' thereby means, that this holds for all inequalities
of the form in Eq.(\ref{FourierBell}). In other words, there
exists a local hidden variable model if we only consider full
correlation functions of two dichotomic observables per site.
Since no counterexample could be found for other cases so far one
might follow Asher Peres\cite{Peresconjecture} and conjecture that
positivity of the partial transpose generally implies the
existence of a local hidden variable model.

Other open problems are associated with the distillability of
entangled states, i.e., the possibility of extracting maximally
entangled states by means of classical communication and local
operations on several copies of the input state. It is for
instance still an open question whether positivity of the partial
transpose is necessary for undistillability\cite{NPT}. Moreover,
it is yet not clear whether the violation of a bipartite Bell
inequality already implies distillability. For multipartite
systems, however, the structure of the state space with respect to
entanglement properties is much richer and D\"ur\cite{DuerBell}
recently showed that there exist indeed undistillable multipartite
states\footnote{The example in Ref.\cite{DuerBell} makes use of at
least eight parties. However, the same techniques can be applied
for 4-partite systems.} violating Mermin's inequality.

Of course Fig. \ref{fig1} is far from being complete. Relations
between Bell inequalities and usefulness for {\it
teleportation}\cite{HHHteleportation}, {\it quantum key
distribution} and {\it quantum secret sharing}\cite{SG1,SG2} have
also been studied. Scarani and Gisin\cite{SG2} for instance have
recently argued that in a secret sharing protocol the authorized
partners have a higher mutual information than the unauthorized
ones, iff they could violate Mermin's inequality. Such a
connection has already been suggested by the coincidence that for
two-partner key distribution a secret key can be established using
one-way privacy amplification iff the two partners can violate the
CHSH inequality \cite{HG97,FGGNP97}.

In this way Bell inequalities seem to appear in a new context,
playing the role of witnesses for the usefulness of a state for
certain quantum communication purposes.

However, the resource point of view of entanglement also requires
a quantitative description, which tells us how much entanglement
is present in a given quantum state. So why not take the maximal
violation of a Bell inequality as a measure of entanglement?
Although, this seems to be quite reasonable at first, the works of
Popescu\cite{Popescu} and Gisin\cite{Gisinexp} have shown that the
maximal violation is in fact not capable of measuring
entanglement. By definition entanglement is that part of
correlations between several subsystems, which is not
``classical''. Therefore a measure of entanglement has to be able
to distinguish this non-classical part from classical correlation,
which can increase under local operations and classical
communication (LOCC). However, Ref.\cite{Popescu,Gisinexp} have
shown in an impressive way, that the maximal violation of a Bell
inequality does not behave monotonously under LOCC operations.
Therefore Bell violations merely give a hint for the strength of
entanglement but they do not fulfill the usual requirements for
entanglement measures.

\nonumsection{Acknowledgements}
Funding by the European Union
project EQUIP (contract IST-1999-11053) and financial support from
the DFG (Bonn) is gratefully acknowledged.

\nonumsection{References} 

\appendix
\noindent

\section{Bell inequalities and convex geometry}

In this appendix we return to the construction of Bell
inequalities, which we started to discuss in Sec.3., and emphasize
that finding all the Bell inequalities is a special instance of a
standard problem in convex geometry known as the {\it convex hull
problem}. This point of view provides a rather intuitive
geometrical interpretation of Bell inequalities as well as making
contact to one of the oldest fields in mathematics\cite{HCG}.

Let us again consider an $n$-partite system, where each party has
the choice of $m$ $v$-valued observables to be measured. Note that
we choose such a symmetric setting just in order to circumvent
cumbersome notation -- the basic idea, however, is obviously
independent of the type of considered correlations.

Hence, we have $m^n$ different experimental setups and each of
them may lead to $v^n$ different outcomes, such that the raw
experimental data are made up of $(mv)^n$ probabilities. These
numbers form a vector $\xi$ lying in a space of dimension $(mv)^n$
(minus a few for normalization constraints), to which we will
refer to as the {\it correlation space}.

Now, we ask for the region $\Xi$ in this correlation space, which
is accessible within any local classical model. The crucial
characteristic of these models is that any vector $\xi$ is
generated by specifying probabilities for each {\it classical
configuration}, i.e., for every assignment of one of the $v$
values to each of the $nm$ observables. Locality is thereby
expressed in the fact that the assignment of a value to an
observable at one site does not depend on the choice of
observables at the other sites.

Since every classical configuration $c$ is also represented by a
vector $\epsilon_c$ of probabilities, the classical accessible
region is just the convex hull of at most $v^{(nm)}$ explicitly
known extreme points:
\begin{equation}\Xi=\mbox{conv}\{\epsilon_c\}.\end{equation}
Since the numbers of configurations is finite $\Xi$ is a convex
{\it polytope}.

\subsection{Representations of convex polytopes and the hull
problem}

Every polytope has two representations. It can either be expressed
in terms of a finite number of extreme points ({\it V
representation}) or as the intersection of halfspaces ({\it H
representation}), i.e., as a set of solutions to a system of
linear inequalities -- which in our case are the Bell
inequalities. The set of linear inequalities corresponding to all
halfspaces containing $\Xi$ is represented by the set
\begin{equation}\label{Polar}
\Xi^*=\{\beta|\forall c : \langle\beta,\epsilon_c\rangle\leq 1\},
\end{equation}
called the {\it polar} of $\Xi$, which is in turn a convex
polytope of the same dimension. The duality between a convex set
and its polar is a generalization of the duality between regular
platonic solids, under which the octahedron and the cube as well
as the dodecahedron and the icosahedron are polars of each other.

It is obvious from the convexity of $\Xi$ that for each
$\beta\in\Xi^*$ the inequality $\langle\beta,\xi\rangle\leq 1$ is
necessary for $\xi\in\Xi$. Moreover, the {\it Bipolar theorem}
\cite{Schaefer} says that the collection of all these inequalities
is also sufficient and since $\Xi^*$ is also convex it suffices to
look at the extreme points of $\Xi^*$. We are therefore left with
the following problem\footnote{See also the {\it problem page}
{\tt http://www.imaph.tu-bs.de/qi/problems/1.html}.} known as the
{\it convex hull} or {\it face enumeration problem}: given the
extreme points of a polytope find the extreme points of its polar.
If there are initially more points than extreme points, then the
convex hull problem is said to be {\it degenerate}.

\subsection{The number of facets}

The total number of Bell inequalities for a given class of
correlations is not known except for the case described in Sec.3.
However, convex geometry provides a number of general results and
upper bounds, which we will just briefly discuss. The most
well-known result in the combinatorial theory of polytopes is
probably the {\it Euler-Poincar$\acute{\mbox{e}}$}\cite{convexity}
relation stating that the numbers $\{f_j\}$ of faces\footnote{A
subset of the polytope $P$ is called a {\it face} if it is the
intersection of the polytope with one of its {\it supporting
hyperplanes}, i.e., a plane $h$ such that one of the closed
halfspaces of $h$ contains $P$. The faces of dimension $0,1,D-1$
are called {\it vertices (extreme points), edges} and {\it
facets}. Every face is again a convex polytope.} of dimension $j$
are related via $\sum_{j=0}^{D-1}(-1)^jf_j=1-(-1)^D$ , where $D$
is the dimension of the polytope. Moreover,
McMullen's\cite{McMullen} upper bound theorem implies that a
polytope with $N$ vertices has at most $f_{D-1}\sim O(N^{\lfloor
D/2 \rfloor})$ facets, which is in our case the number of Bell
inequalities. This bound is also tight as exhibited by {\it cyclic
polytopes}. However, the class of polytopes occurring in the
construction of Bell inequalities is of a rather special type
often referred to as {\it 0-1 polytopes} since the components of
any extreme point $\epsilon_c$ are either 0 or 1. The question
whether the number of facets of such polytopes is bounded by an
exponential in $D$ was just recently given a negative answer to by
B\'ar\'any and P\'or \cite{BaranyPor}, who showed that there
exists a positive constant $c$ such that
\begin{equation}\label{BPEq}
f_{D-1}>\Big(\frac{c D}{\log D}\Big)^{D/4},
\end{equation}
for some 0-1 polytopes. Hence the growth can in general be
superexponential.

\subsection{Complexity of convex hull algorithms}

There are several ways of measuring the complexity of a convex
hull algorithm\cite{chullalgorithms}. Basically, there are two
points in which different approaches differ from each other: the
elementary operations (bit operations vs. elementary arithmetic
operations) and the role of the output (whether it is a parameter
of the measure or not).

One way would be just to count the number of elementary arithmetic
operations and to assume that the storage of an integer number
takes a unit space (which is essentially the {\it unit cost RAM
model} as opposed to the {\it Turing model}). Fixing the dimension
$D$ of the polytope and then looking at the worst-case running
time as a function of $N$ an optimal algorithm is
known\cite{Cha93} that runs in time $O(N^{\lfloor D/2 \rfloor})$,
as already suggested by McMullen's upper bound theorem.

For the general nondegenerate convex hull problem there are
algorithms with run times which are polynomially bounded by $N,D$
and $f_{D-1}$ (e.g.\cite{AF92}). For the degenerate case however
no such polynomial algorithm is known.

For a more detailed discussion of the complexity of finding a
complete set of Bell inequalities we  would like to refer to the
work of Pitowsky\cite{Pitowskybook,Pcomplex}, who also discusses
the relation between the convex hull problem and the notorious $NP
= P$ resp. $NP = coNP$ questions. It is shown, that deciding
membership in a ``correlation polytope'' is an NP-complete
problem, whereas deciding facets is probably not even in NP.
Moreover, in Ref.\cite{Pcomplex} the relation to the minimum
energy problem for Ising spin systems is discussed, which in turn
was shown to be NP-hard by Barahona\cite{Barahona}.


\noindent

\begin{thebibliography}{000}

\bibitem{EPR} A. Einstein, B. Podolsky, and N. Rosen, Phys. Rev.
{\bf 47} , 777 (1935).
\bibitem{schroed} E. Schr\"odinger, Naturwissenschaften {\bf 23},
807-812; 823-828; 844-849 (1935).
\bibitem{Bohr} N. Bohr, Nature {\bf 136}, 65 (1935); Phys. Rev.
{\bf 48}, 696 (1935).
\bibitem{Bohm1} D. Bohm, {\sl Quantum Theory} (Prentice-Hall,
Englewood Cliffs, New York, 1951).
\bibitem{Bell64} J.S. Bell, Physics {\bf 1}, 195 (1964).
\bibitem{Aspect} A. Aspect, P. Grangier, and G. Roger, Phys. Rev.
Lett. {\bf 47}, 460 (1981).
\bibitem{detloop} M. A. Rowe et al., Nature {\bf 409}, 791 (2001).
\bibitem{Werner89} R.F. Werner, Phys. Rev. A {\bf 40}, 4277 (1989).
\bibitem{Popescu} S. Popescu, Phys. Rev. Lett. {\bf 74}, 2619 (1995).
\bibitem{B96} C.H. Bennett, G. Brassard, S. Popescu, B.
Schumacher, J. Smolin, and W.K. Wootters, Phys. Rev. Lett. {\bf
76}, 722 (1996).
\bibitem{Gisinexp} N. Gisin, Phys. Lett. A {\bf 210}, 151 (1996).
\bibitem{BE}  M. Horodecki, P. Horodecki, and R. Horodecki, Phys. Rev. Lett. {\bf
 80}, 5239 (1998).
\bibitem{PPTBell} R.F. Werner and M.M. Wolf, Phys. Rev. A {\bf
61}, 062102 (2000).
\bibitem{WW2} R.F. Werner and M.M. Wolf, quant-ph/0102024 (2001).
\bibitem{DuerBell} W. D\"ur, quant-ph/0107050 (2001).
\bibitem{AC} A. Cabello, Phys. Rev. Lett. {\bf 86}, 1911 (2001).
\bibitem{CSreview} J.F. Clauser and A. Shimony, Rep. Prog. Phys.
{\bf 41}, 1881 (1978).
\bibitem{HSreview} D. Home and F. Selleri, Riv. Nuovo Cimento {\bf
14}, 1 (1991).
\bibitem{higherspins} A. Garg and N.D. Mermin, Found. Phys. {\bf
14}, 1 (1984).
\bibitem{Merminrev} N.D. Mermin, Rev. Mod. Phys. {\bf 65}, 803
(1993).
\bibitem{vNeuHidd} J.v. Neumann, {\it Mathematische Grundlagen der
Quantenmechanik}, (Springer, Berlin, 1932).

\bibitem{Bohm2} D. Bohm, Phys. Rev. {\bf 85}, 166 (1952); {\bf
85}, 180 (1952).
\bibitem{Nelson} E. Nelson, {\it Dynamical Theories of Brownian
Motion}, (Princeton University Press 1967).
\bibitem{WerHidden}R.F. Werner, Phys. Rev D {\bf 34}, 463 (1986).

\bibitem{Fine} A. Fine,  Phys. Rev. Lett. {\bf 48}, 291 (1982); A.
Fine, J. Math. Phys. {\bf 23}, 1306 (1982).

\bibitem{CHSH} J. F. Clauser, M. A. Horne, A. Shimony, and
R. A. Holt, Phys. Rev. Lett. {\bf 23}, 880 (1969).

\bibitem{invitation} R.F. Werner in {\it Quantum Information -- an
introduction to basic theoretical concepts and experiments}
(Springer tracts in modern physics, Berlin, 2001).

\bibitem{Pitowskybook} I. Pitowsky, {\it Quantum Probability -- Quantum Logic} (Springer, Berlin, 1989).
\bibitem{Pcomplex} I. Pitowsky, Mathematical Programming {\bf 50}, 395
(1991).
\bibitem{ZukowskiBrukner} M. Zukowski and C. Brukner,
quant-ph/0102039 (2001).
\bibitem{PitowskySvozil} I. Pitowsky and K. Svozil, quant-ph/0011060 (2000).
\bibitem{PSweb} A list of the coefficients belonging to the 53856 inequalities for the case $(n,m,v)=(3,2,2)$ can be found on {\tt
http://tph.tuwien.ac.at/~svozil/publ/ghzbig.html}.
\bibitem{nsettings} N. Gisin, Phys. Lett. A {\bf 260}, 1 (1999).
\bibitem{MerminGHZ} N.D. Mermin, Phys. Rev. Lett. {\bf 65}, 3373
(1990).
\bibitem{Kaszlikowski} D. Kaszlikowski, P. Gnacinski, M. Zukowski,
W. Miklaszewski, and A. Zeilinger, Phys. Rev. Lett. {\bf 85}, 4418
(2000).
\bibitem{GargMerminvsFine} A. Garg and N.D. Mermin, Phys. Rev.
Lett. {\bf 49}, 1220 (1982).
\bibitem{Merminineq}  N. D. Mermin, Phys. Rev. Lett. {\bf 65}, 1838 (1990).
\bibitem{Ardehali} M. Ardehali, Phys. Rev. A {\bf 46}, 5375 (1992).
\bibitem{Roy} S. M. Roy and V. Singh, Phys. Rev. Lett. {\bf 67}, 2761 (1991).
\bibitem{Klyshko} A. V. Belinskii and D. N. Klyshko, Sov. Phys. Usp. {\bf 36}, 653 (1993).
\bibitem{Gisin} N. Gisin and H. Bechmann-Pasquinucci, Phys. Lett. A {\bf 246}, 1 (1998).
\bibitem{Peresconjecture} A. Peres, Found. Phys. {\bf 29}, 589 (1999).
\bibitem{PeresPPT} A. Peres, Phys. Rev. Lett. {\bf 77}, 1413 (1996).

\bibitem{dis} C. H. Bennett, G. Brassard, S. Popescu, B. Schumacher,
    J. A. Smolin, and W. K. Wootters, Phys. Rev. Lett. {\bf 76}, 722 (1996).
\bibitem{Cirelsonbound} B. S. Cirel'son, Lett. Math. Phys. {\bf 4}, 93 (1980).
\bibitem{Landau} L. J. Landau, Phys. Lett. A {\bf 120}, 54   (1987).
\bibitem{SW2}S.J. Summers and  R.F. Werner, Commun. Math. Phys. {\bf 110}, 247 (1987).
\bibitem{SWPauli} S.J. Summers and R.F. Werner, Ann. Inst. Henri
Poincar\'e {\bf 49}, 215 (1988).
\bibitem{Gisindecomposition} V. Scarani and N. Gisin,
quant-ph/0103068 (2001).
\bibitem{HHHCHSH} R. Horodecki, P. Horodecki, and M. Horodecki,
Phys. Lett. A {\bf 200}, 340 (1995).
\bibitem{Gisinpure} N. Gisin, Phys. Lett. A {\bf 154}, 201 (1991).
\bibitem{GisinPeres} N. Gisin and A. Peres, Phys. Lett. A
{\bf 162}, 15 (1992).
\bibitem{BW98} K. Banaszek and K. W$\acute{\mbox{o}}$dkiewicz, Phys. Rev. A {\bf
58}, 4345 (1998).
\bibitem{BW99} K. Banaszek and K. W$\acute{\mbox{o}}$dkiewicz, Phys.
Rev. Lett. {\bf 82}, 2009 (1999).
\bibitem{JLK00} H. Jeong, J. Lee, M.S. Kim, Phys. Rev. A {\bf 61},
052101 (2000).
\bibitem{CPHZ01} Z.-B. Chen, J.-W. Pan, G. Hou, and Y.-D. Zhang,
quant-ph/0103051 (2001).
\bibitem{HaagKastler} R. Haag and D. Kastler, J. Math. Phys. {\bf 5}, 848 (1964).
\bibitem{Haag} R. Haag, {\it Local Quantum Physics}, Springer (1992).
\bibitem{SW1} S.J. Summers and R.F. Werner, Phys. Lett. A {\bf 110}, 257 (1985).
\bibitem{VerchW} R. Verch and R.F. Werner, in preparation.
\bibitem{BraunsteinCaves} S.L. Braunstein and C.M. Caves, Ann.
Phys. {\bf 202}, 22 (1990).
\bibitem{CC1} B.S. Cirel'son, Lett. Math. Phys., {\bf 4}, 93
(1980).
\bibitem{CC2} B.S. Tsirel'son, J. Sov. Math., {\bf 36}, 557 (1987).
\bibitem{GHZ89} D.M. Greenberger, M. Horne, and A. Zeilinger in
{\it Bell's Theorem, Quantum Theory, and Conceptions of the
Universe}, M. Kafatos, ed., Kluwer, Dordrecht (1989).
\bibitem{Zukowskiallsettings} M. Zukowski, Phys. Lett. A {\bf 177}, 290
(1993).
\bibitem{NPT} D.P. DiVincenzo, P.W. Shor, J.A. Smolin, B.M. Terhal, and A.V.
     Thapliyal,  Phys. Rev. A {\bf 61}, 062312 (2000);  W. D\"ur, J.I. Cirac, M. Lewenstein, D. Bruss, Phys. Rev. A {\bf 61}, 062313
(2000); T. Eggeling, K.G.H. Vollbrecht, R.F. Werner, and M.M.
Wolf, quant-ph/0104095 (2001).
\bibitem{HHHteleportation} R. Horodecki, M. Horodecki, and P.
Horodecki, quant-ph/9606027 (1996).
\bibitem{SG1} V. Scarani and N. Gisin, quant-ph/0101110 (2001).
\bibitem{SG2} V. Scarani and N. Gisin, quant-ph/0104016 (2001).
\bibitem{HG97} B. Huttner and N. Gisin, Phys. Lett. A {\bf 228},
13 (1997).
\bibitem{FGGNP97} C. Fuchs, N. Gisin, R.B. Griffiths, C.-S. Niu,
and A. Peres, Phys. Rev. A {\bf 56}, 1163 (1997).
\bibitem{HCG} P.M. Gruber and  J.M. Wills (editors), {\it Handbook of convex geometry},
North-Holland (1993).
\bibitem{Schaefer} H.H. Schaefer, {\it Topological Vector Spaces} (Springer, Berlin, 1980).
\bibitem{convexity} R. Webster, {\it Convexity} (Oxford University Press,
1994).
\bibitem{McMullen} P. McMullen, J. Combin. Theory Ser. B {\bf 10},
187 (1971).
\bibitem{BaranyPor} I. B\'ar\'any and A. P\'or, {\it 0-1 polytopes with many
facets}, manuscript, R\'enyi Institute of Mathematics, Hungarian
Academy of Sciences; {\tt http://www.renyi.hu/~barany/} (2000).
\bibitem{chullalgorithms} C.K. Yap, {\it Fundamental problems in
algorithmic algebra} (Oxford University Press, 2000); see also
{\tt http://www.ifor.math.ethz.ch/fukuda/polyfaq/ \\ polyfaq.html}
for a brief review.
\bibitem{Cha93} B. Chazelle, Discrete Compute. Geom. {\bf 10}, 377
(1993).
\bibitem{AF92} D. Avis and K. Fukuda, Discrete Compute. Geom. {\bf
8}, 295 (1992).
\bibitem{Barahona} F. Barahona, J. Phys. A {\bf 15}, 3241 (1982).
\end{thebibliography}
\end{document}